\documentclass[showpacs,twocolumn,aps,amsmath]{revtex4}

\usepackage{graphicx}

\begin{document}

\title{Compact quantum gates on electron-spin qubits assisted by  diamond nitrogen-vacancy centers inside
cavities\footnote{Published in Phys. Rev. A \textbf{88}, 042323
(2013)}}

\author{Hai-Rui Wei and Fu-Guo Deng\footnote{Corresponding author: fgdeng@bnu.edu.cn}}

\address{Department of Physics, Applied Optics Beijing Area Major Laboratory, Beijing Normal University, Beijing 100875, China}

\date{\today }

\begin{abstract}
Constructing  compact quantum circuits for universal quantum gates
on solid-state systems is crucial for quantum computing.  We present
some compact quantum circuits for a deterministic solid-state
quantum computing, including the CNOT, Toffoli, and Fredkin gates on
the diamond nitrogen-vacancy centers confined inside cavities,
achieved by some input-output processes of a single photon. Our
quantum circuits for these universal quantum gates are simple and
economic. Moreover, additional electron qubits are not employed, but
only a single-photon medium. These gates have a long coherent time.
We discuss the feasibility of these universal solid-state quantum
gates, concluding that they are feasible with current technology.
\end{abstract}

 \maketitle

\section{Introduction}\label{sec1}

Quantum logic gates are the key elements in quantum computing.  It
is well known that two-qubit entangling gates can be used to
implement any $n$-qubit quantum computing,  assisted by single-qubit
gates \cite{book,universal}. The family composed of controlled-NOT
(CNOT) gates and one-qubit gates is the most popular universal set
of quantum gates for quantum computing today
\cite{cnota1,cnota2,cnota3,CNOT2,cnota4,weioe,RenLPL,longpra,longprl,cnota5,CNOT-Hybrid,weipra,CNOT1}.
The simulation of any two-qubit gate requires at least three CNOT
gates and 15 single-qubit rotations
\cite{3CNOT1,3CNOT2,3CNOT3,3CNOT4,small}. Therefore, projects for
realizing a CNOT gate  in a solid-state system are highly desired
for  quantum computing in the future.

An optimal unstructured quantum circuit for any multi-qubit gate
requires $[\frac{1}{4}(4^n-3n-1)]$ CNOT gates \cite{3CNOT4}. In the
domain of a three-qubit case,  people  pay  much  attention   to
Toffoli \cite{Toffoli} and Fredkin gates \cite{Fredkin}. \{Toffoli
(Fredkin) gate, Hadamard gates\} is a universal set for multi-qubit
quantum computing \cite{Toffoli,Fredkin}. It is usual much more
complex and difficult to realize a Toffoli gate or a Fredkin gate
with CNOT and one-qubit gates in experiment because it requires at
least six CNOT gates \cite{Toffolicost} to  synthesize  a Toffoli
gate and it requires two CNOT and three
controlled-$\sqrt{\text{NOT}}$ gates \cite{Fredkincost} to
synthesize a Fredkin gate. It is particularly interesting to discuss
the physical realization of  a Toffoli gate and a Fredkin gate in a
simpler way.

Quantum gates on solid-state systems have attracted much attention
as they have a good scalability, and it has been demonstrated for
superconducting qubits \cite{superPRL,superPRB,super} and quantum
dots \cite{QD}. Electron-spin qubits in solid-state systems, in
particular, associated with nitrogen-vacancy (NV) defect centers,
are particularly attractive.

The negatively charged NV defect center occurs in the diamond
lattice consisting of a substitutional $^{14}$N atom and an adjacent
vacancy, and is one of the most attracting and promising solid-state
candidates for quantum information processing, due to the long
room-temperature coherent time (1.8 $ms$) \cite{coherence1} that can
be manipulated and coupled together in a scalable fashion. The
procedures have been established for optical initialing, optical
preparing, fast microwave or magnetic manipulating, and optical
detecting the long-lived spin triplet state associated with NV
centers
\cite{manipulate2,manipulate3,manipulate4,manipulate5,manipulate6,ODMR2}.

Tremendous theoretical and experimental progress has been made on
quantum information processing based on NV centers. The schemes for
the quantum entanglement generation between a photon and an NV
center \cite{photon-NV}, and between electrons associated with NV
 centers \cite{NV-NV1,NV-NV2,NV-NV3,NV-NV4,NV-NV5} were proposed. Recently, the schemes for the quantum state transfer
between separated  NV centers were introduced
\cite{transfer1,transfer2,transfer3}. Multiqubit quantum registers
associated with separated NV centers in diamonds have been proposed
 \cite{NV-NV1,NV-NV2,transfer1}. Hyperentanglement purification and concentration of two-photon systems in both the spatial-mode
and polarization degrees of freedom were investigated
\cite{renhyperepp} with the assistance of diamond NV centers inside
photonic crystal cavities.   Yang \emph{et al.} \cite{CCPF} proposed
a scheme for implementing the conditional phase gate between NV
centers assisted by a high-Q silica microsphere cavity. As the
electron spin of the NV defect center couples to nearby $^{13}$C
nuclear spins, a high-fidelity polarization and the detection of the
single-electron and nuclear-spin states can be achieved, even under
ambient conditions
\cite{detect-nuclear2,detect-nuclear3,detect-nuclear4,detect-nuclear5},
which  allows  quantum information transfer
\cite{register1,register2,nuclear-nuclear}, entanglement generation
between an electron-spin qubit and a nuclear-spin qubit
 \cite{electron-nuclear,elec-register}  and between two nuclear spins
 \cite{nuclear-nuclear}, and the construction of the quantum gate between an electron
and a nuclear spin \cite{CROT}.

In 2011, Chen \emph{et al.} \cite{NV-NV3}  proposed a composite
system, i.e., a diamond NV$^-$ center with six electrons from the
nitrogen and three carbons surrounding the vacancy, which is
confined in a microtoroidal resonator (MTR) \cite{MTR} with a
quantized whispering-gallery mode (WGM). This system allows for an
ultrahigh-$Q$ and a small mode volume of WGM microresonators
\cite{ultrahigh1,ultrahigh2,ultrahigh3}. When the MTR couples to the
fiber, the ultrahigh-Q is degraded. The experiments in which a
diamond NV center couples to WGMs in a silica microsphere
\cite{microsphere1,microsphere2,microsphere3}, diamond-GaP microdisk
\cite{microdisk}, or SiN photonic crystal \cite{crystal} have been
demonstrated. The photon input-output process of a coupled
 atom  and MTR platform has been demonstrated in experiment
\cite{MTR}.

It is important to construct compact quantum circuits for universal
quantum gates  because they  reduce not only time but also errors.
In this paper, we investigate the possibility of constructing
compact universal quantum gates for a  deterministic solid-state
quantum computing, including the CNOT, Toffoli, and Fredkin gates on
the diamond NV centers confined in cavities, by some single-photon
input-output processes. The qubits of these deterministic gates are
encoded on two of the electron-spin triple ground states associated
with the diamond NV centers, and they have a long decoherence time
even at the room temperature. Our quantum gates on NV centers are
obtained by interacting a photon with the NV centers, detecting the
emitting photon medium, and applying some proper feedforward
operations on the electron-spin qubits associated with NV centers.
Our quantum circuits for these gates are compact and economic. The
CNOT and Toffoli gates are particularly appealed as the photon
medium only interacts with each electron qubit one time. Compared
with the synthesis programs, our schemes are simple. In our
proposals, auxiliary electron-spin qubits are not required and only
one photon medium is employed, which is different from the quantum
gates on moving electrons based on charge detection \cite{CNOT1} and
the photonic quantum gates based on cross-Kerr nonlinearities
\cite{CNOT2}. With current technology, these universal solid-state
quantum gates are feasible. If the photon loss, the detection
inefficiency, and the imperfection of the experiment are negligible,
the success probabilities of our gates are 100\%.

This article is organized as follows. In Sec. \ref{sec2}, we
introduce the photon-matter platform based on the diamond NV center
coupled to a resonator and the compact quantum circuit for  a
deterministic CNOT gate on two separated diamond NV centers.
Subsequently, the quantum circuits for constructing three-qubit
Toffoli and  Fredkin gates on three separated diamond NV centers in
a deterministic way  are given in Secs. \ref{sec3} and \ref{sec4},
respectively.  The fidelities and efficiencies of our proposals are
estimated in Sec. \ref{sec5}. Finally, we discuss the feasibility of
our universal quantum gates and give a summary  in Sec. \ref{sec6}.

\begin{figure}[tpb]           
\begin{center}
\includegraphics[width=6.0 cm,angle=0]{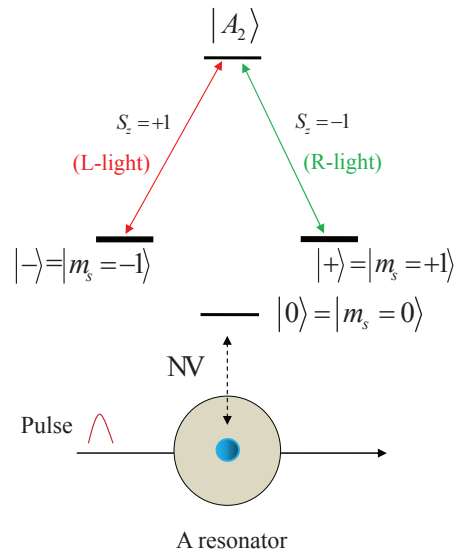}
\caption{(Color online) Schematic diagram of a diamond NV center
coupling to a  resonator and the possible $\Lambda$-type optical
transitions in an NV center. The transition
$|-\rangle\rightarrow|A_2\rangle$ is derived by a left-circularly
polarized photon (denoted by $\vert L\rangle$ or $ S_{z}=+1 $), and
$|+\rangle\rightarrow|A_2\rangle$ is derived by a right-circularly
polarized photon (denoted by $\vert L\rangle$ or $ S_{z}=-1 $). The
levels in bold encode the qubits, i.e., $|+\rangle=|m_s=+1\rangle$
and $|-\rangle=|m_s=-1\rangle$.} \label{Fig1}
\end{center}
\end{figure}

\section{Two-qubit controlled-not gate on an NV-center system} \label{sec2}

\subsection{A diamond NV center coupled to an MTR with a WGM} \label{sec21}

The electron-spin triple ground states of an  NV center are split
into $|m_s=0\rangle$ (denoted by $|0\rangle$) and $|m_s=\pm1\rangle$
(denoted by $|\pm\rangle$) by 2.88 GHz with zero-field, due to the
spin-spin interactions \cite{split}. The structure of the excited
states is relatively complex, and it includes six excited states
defined by the method of group theory \cite{photon-NV},
$|A_1\rangle=(|E_-\rangle|+\rangle-|E_+\rangle|-\rangle)/\sqrt{2}$,
$|A_2\rangle=(|E_-\rangle|+\rangle+|E_+\rangle|-\rangle)/\sqrt{2}$,
$|E_x\rangle=|X\rangle|0\rangle$, $|E_y\rangle=|Y\rangle|0\rangle$,
$|E_1\rangle=(|E_-\rangle|-\rangle-|E_+\rangle|+\rangle)/\sqrt{2}$,
and
$|E_2\rangle=(|E_-\rangle|-\rangle+|E_+\rangle|+\rangle)/\sqrt{2}$,
owing to NV center's C$_{3v}$ symmetry, spin-spin, and spin-orbit
interactions in the absence of external magnetic field or crystal
strain. Here, $|E_\pm\rangle$, $
 |X\rangle=(|E_{-}\rangle-|E_{+}\rangle)/2$ and $|Y\rangle=i(|E_{-}\rangle+|E_{+}\rangle)/2$ are the orbital states,
 and  $|E_\pm\rangle$ has angular momentum projections $\pm1$  along the NV axis.

In our work,  the quantum information of the quantum gate is encoded
on the spins of the electronic ground triple states
$|+\rangle=|m_s=1\rangle$ and $|-\rangle=|m_s=-1\rangle$. The
$\Lambda$-type three-level system (see Fig.\ref{Fig1}) is realized
by employing one of the specific excited state $|A_2\rangle$ as an
ancillary state \cite{photon-NV}. The $\Lambda$-type system in which
optical control is required, can be obtained by using a particular
magnetic field to mix the ground states \cite{magnetic}.
Alternatively, it is possible to find a $\Lambda$-type system  at
zero magnetic field as the inevitable strain in diamond reduces the
symmetry and primarily  modifies  the  excited-state structure
according to their orbital wave functions. The excited state is
separated into two branches \cite{strain1,strain2}, $|A_{1}\rangle$,
$|A_{2}\rangle$, $|E_x\rangle$, and $|E_y\rangle$, $|E_{1}\rangle$,
$|E_{2}\rangle$ at  moderate and high strain. Togan \emph{et al.}
\cite{photon-NV} demonstrated that the state $|A_2\rangle$ is robust
to low strain and magnetic fields due to the stable symmetric
properties, and it decays with an equal probability to the
ground-state sublevels $|-\rangle$ through a left circularly
polarized radiation $|L\rangle$ ($ S_{z}=+1 $) and to $|+\rangle$
through a right circularly polarized radiation $|R\rangle$  ($
S_{z}=-1 $). That is, the zero phonon line (ZPL) was observed after
the optical resonant excitation at 637 $nm$
($|-\rangle\rightarrow|A_2\rangle$ driven by a $L$-polarized photon
and $|+\rangle\rightarrow|A_2\rangle$ driven by a $R$-polarized
photon). The mutually orthogonal circular polarization will be
destroyed by high strain. The preparation and measurement of the
electron spin can be realized by exploiting resonant optical
excitation techniques. As illustrated in Ref. \cite{photon-NV}, the
electron spin can be polarized by first preparing the electron spin
to $|0\rangle$ by means of optical pumping with a 532-$nm$ light,
and then transferring the population to either $|\pm\rangle$ by
means of microwave $\pi$ pulses. The spin can be a high-fidelity
($\sim$93.2\%) readout and addressed at low temperature (T=8.6K)
based on spin-dependent optical transitions. The state $|A_2\rangle$
connects $|\pm 1\rangle$, and $|E_{x,y}\rangle$ connects
$|0\rangle$, after spin manipulation by a microwave pulse and
resonant excitation transition
$|0\rangle\leftrightarrow|E_{x,y}\rangle$. The presence or absence
of fluorescence decay reveals the spin state
\cite{photon-NV,elec-register}.

The Heisenberg equations of the motion for the annihilation operator
of the cavity mode $\hat{a}$ and the lowing operator of the NV
center operation $\sigma_-$ and the input-output relation for the
cavity are given by \cite{QObook}
\begin{eqnarray}       \label{eq1}
\frac{d\hat{a}}{dt}&=& -\left[i(\omega_c-\omega_p)+\frac{\kappa}{2}\right]\hat{a}(t)-g\sigma_{-}(t) - \sqrt{\kappa}\,\hat{a}_{in}, \nonumber\\
\frac{d\sigma_-}{dt}&=& -\left[i(\omega_{0}-\omega_p)+\frac{\gamma}{2}\right]\sigma_{-}(t)-g\sigma_z(t)\;\hat{a}(t)\nonumber\\
&& + \sqrt{\gamma}\,\sigma_z(t)\;\hat{b}_{in}(t), \nonumber\\
\hat{a}_{out} &=& \hat{a}_{in}+ \sqrt{\kappa}\;\hat{a}(t),
\end{eqnarray}
where $\omega_c$, $\omega_p$, and $\omega_0$ are the frequencies of
the cavity, the single photon, and the  NV center, respectively.
$\hat{a}_{in}(t)$ and $\hat{a}_{out}$  are the cavity input and
output operators, respectively. $\sigma_z(t)$ is the inversion
operator of the cavity. $\gamma$ is the decay of the NV center.
$\kappa$ is the damping rate of the cavity. $g$ is the coupling
rate. $b_{in}(t)$ is the vacuum input field felt by the NV center
with the commutation relation
$[\hat{b}_{in}(t),\hat{b}_{in}^\dag(t')]=\delta(t-t')$.

In a weak excitation, i.e., taking $\langle \sigma_z\rangle= -1$,
the adiabatical elimination of the cavity mode leads to the
reflection coefficient of the NV center confined in the
 cavity as \cite{Hersenberg,Hu}
\begin{eqnarray}       \label{eq2}
r(\omega_p) &=&\frac{\hat{a}_{out}}{\hat{a}_{in}}
=\frac{[i(\omega_{c}-\omega_p)-\frac{\kappa}{2}][i(\omega_{0}-\omega_p)+\frac{\gamma}{2}]+g^2}
           {[i(\omega_{c}-\omega_p)+\frac{\kappa}{2}][i(\omega_{0}-\omega_p)+\frac{\gamma}{2}]+g^2}.\nonumber\\
\end{eqnarray}

The phase shift and the amplitude of the reflected photon are a
function of the frequency detuning $\omega_c-\omega_p$, with
$\omega_c=\omega_0$.  For $\omega_c=\omega_0=\omega_p$, i.e., when
the cavity mode  resonant with the NV center interacts with the
resonant photon pulse, one can obtain  \cite{Hersenberg}
\begin{eqnarray}       \label{eq3}
r(\omega_p)=\frac{-\frac{\kappa\gamma}{4}+g^2}{\frac{\kappa\gamma}{4}+g^2},
\quad\quad\quad
 r_0(\omega_p)=-1.
\end{eqnarray}
Here, $r_0$ is the reflection coefficient of the cold (or the empty)
cavity, that is, $g=0$ and the cavity is not coupled to the NV
center. $r(\omega_p)$ is the one for  the hot cavity, i.e.,
$g\neq0$. Therefore, the change of the input photon is summarized as
 \cite{NV-NV3}
\begin{eqnarray}       \label{eq4}
|R\rangle|+\rangle  &\;\rightarrow\;&  r|R\rangle|+\rangle, \nonumber\\
|L\rangle|-\rangle  &\;\rightarrow\;&  r|L\rangle|-\rangle, \nonumber\\
|R\rangle|-\rangle  &\;\rightarrow\;&  -|R\rangle|-\rangle, \nonumber\\
|L\rangle|+\rangle  &\;\rightarrow\;&  -|L\rangle|+\rangle.
\end{eqnarray}
The effect of the coupling strength  $g/\sqrt{\kappa\gamma}$ on the
amplitude of the reflected  photon and that of the frequency
detuning on the phase shift have been discussed in \cite{NV-NV3}.
Chen \emph{et al.} \cite{NV-NV3} showed that when
$g\geq5\sqrt{\gamma\kappa}$ with $\omega_c=\omega_0=\omega_p$,
\begin{eqnarray}       \label{eq5}
r(\omega_p)\simeq1, \quad \quad \quad
 r_0(\omega_p) =-1.
\end{eqnarray}
That is, Eq. (\ref{eq4}) becomes
\begin{eqnarray}       \label{eq6}
|R\rangle|+\rangle &\;\rightarrow\;&  |R\rangle|+\rangle,\nonumber\\
|L\rangle|-\rangle &\;\rightarrow\;&  |L\rangle|-\rangle,\nonumber\\
|R\rangle|-\rangle &\;\rightarrow\;&  -|R\rangle|-\rangle, \nonumber\\
|L\rangle|+\rangle &\;\rightarrow\;&  -|L\rangle|+\rangle.
\end{eqnarray}

From the $\Lambda$-type  diamond NV-center optical transition
depicted by Fig. \ref{Fig1}, one can see that it requires a
polarization-degenerate cavity mode. Therefore, it is suitable for
not only WGM microresonators \cite{NV-NV2,NV-NV3,MTR,degenerate2},
but also H1 photonic crystals
\cite{unpolarized-photon1,unpolarized-photon2}, micropillars
\cite{unpolarized-pillar1,unpolarized-pillar2,unpolarized-pillar3},
and fiber-based \cite{fiber-based} cavities.

In our work,  all the devices  work  under the resonant condition
$\omega_{c}=\omega_0=\omega_p$. In the following, we first consider
the case $g\geq5\sqrt{\kappa\gamma}$, that is, $r(\omega_p)\simeq1$,
and then we discuss the effect of $g/\sqrt{\kappa\gamma}$ on the the
fidelities and the efficiencies of our universal quantum gates on
NV-center systems.

\subsection{Compact quantum circuit for  a two-qubit controlled-not gate on an NV-center system}
\label{sec22}

\begin{figure}[!h]          
\begin{center}
\includegraphics[width=8.0 cm,angle=0]{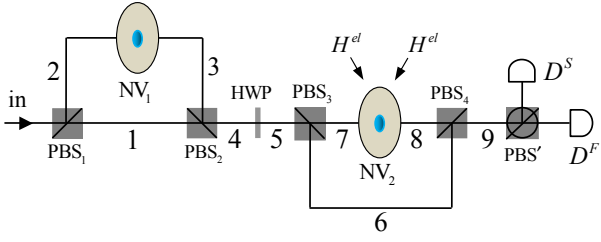}
\caption{(Color online) Compact quantum circuit for  a  CNOT gate on
two NV centers. HWP is a half-wave plate set at 22.5$^\circ$ to
complete the Hadamard operation ($H^{ph}$) on the polarization
photon. The polarizing beam splitter PBS$_i$ ($i=1,2$) in the basis
$\{\vert R\rangle, \vert L\rangle\}$  transmits the right-circularly
polarized photon $|R\rangle$ and reflects the left-circularly
polarized photon $|L\rangle$, respectively. PBS$\,'$ represents a
PBS which transmits the photon in the state
$|F\rangle=(|R\rangle+|L\rangle)/\sqrt{2}$ and reflects the photon
in the state $|S\rangle=(|R\rangle-|L\rangle)/\sqrt{2}$,
respectively. $D^F$ and $D^S$ are two single-photon detectors.}
\label{CNOT}
\end{center}
\end{figure}

Our quantum circuit for  a CNOT gate on two NV centers is shown in
Fig. \ref{CNOT} . The two NV centers are initially prepared in two
arbitrary superpositions of the two ground states $|+\rangle$ and
$|-\rangle$; that is,
\begin{eqnarray}                    \label{eq7}
|\psi\rangle_{c}^{el} &=& \alpha _{c}|+ \rangle_{c}+\beta_c|-\rangle_{c},\nonumber \\
|\psi\rangle_{t}^{el}&=&\alpha_{t}|+\rangle_{t}+\beta_t|-\rangle_{t}.
\end{eqnarray}
Here $|\alpha_c|^2 +|\beta_c|^2 = |\alpha_t|^2 + |\beta_t|^2=1$. The
subscripts $c$ and $t$ stand for the control qubit NV$_1$  and the
target qubit  NV$_2$,  respectively. The single-photon medium is
initially prepared in the equal superposition of $|R\rangle$ and
$|L\rangle$; that is,
\begin{eqnarray}                    \label{eq8}
|\psi_0\rangle^{ph} =\frac{1}{\sqrt{2}}(|R\rangle+|L\rangle).
\end{eqnarray}

Polarizing beam splitter PBS$_1$ splits the input single photon into
two wave-packets. The component $|R\rangle$ transmits through
PBS$_1$ and then arrives at PBS$_2$ directly, while the component
$|L\rangle$ is reflected to spatial model 2 for interacting with
NV$_1$, which induces the transformation
$|L\rangle_2(\alpha_c|+\rangle_c+\beta_c|-\rangle_c)\xrightarrow{\text{NV}_1}|L\rangle_3(-\alpha_c|+\rangle_c+\beta_c|-\rangle_c)$.
Here and the after, the subscript $i$ of $|L\rangle_i$ (or
$|R\rangle_i$, $i=1,2,3,\cdots$) stands for the spatial mode $i$
from where the $L$-polarized photon ($R$-polarized photon) emits.
After the $|R\rangle_1$ and the $|L\rangle_3$ wave packets arrive at
PBS$_2$ simultaneously, the photon emits from spatial mode 4. The
specific evolution process of the whole system composed of the input
photon and two NV centers can be shown as follows:
\begin{eqnarray}                      \label{eq9}
&&|\Psi_{0}\rangle=|\psi_0\rangle^{ph}\otimes
|\psi\rangle_{c}^{el}\otimes|\psi\rangle_{t}^{el}\nonumber\\
&\xrightarrow{\text{PBS}_{1}}&|\Psi_{1}\rangle=\frac{1}{\sqrt{2}}(|R\rangle_1
+|L\rangle_2)\otimes|\psi\rangle_{c}^{el}\otimes|\psi\rangle_{t}^{el}\nonumber\\
&\xrightarrow{\text{NV}_{1}}&|\Psi_{2}\rangle=\frac{1}{\sqrt{2}}|R\rangle_1(\alpha_{c}|
+\rangle_{c}+\beta_c|-\rangle_{c})\otimes|\psi\rangle_{t}^{el}\nonumber\\
                           &&+\frac{1}{\sqrt{2}}|L\rangle_3(-\alpha_{c}|+\rangle_{c}+\beta_c|-\rangle_{c})\otimes|\psi\rangle_{t}^{el}\nonumber\\
&\xrightarrow{\text{PBS}_{2}}& |\Psi_3\rangle=
\frac{1}{\sqrt{2}}|R\rangle_4(\alpha_{c}|+\rangle_{c}+\beta_c|-\rangle_{c})\otimes|\psi\rangle_{t}^{el}\nonumber\\&&
+\frac{1}{\sqrt{2}}|L\rangle_4(-\alpha_{c}|+\rangle_{c}+\beta_c|-\rangle_{c})\otimes|\psi\rangle_{t}^{el}.
\end{eqnarray}
From Eq. (\ref{eq9}), one can see that the balanced Mach-Zehnder
(MZ) interferometer composed of PBS$_1$, NV$_1$, and PBS$_2$
completes the operation
\begin{eqnarray}                       \label{eq10}
\text{PBS}_{1}\rightarrow \text{NV}_{1}\rightarrow
\text{PBS}_{2}=\left(\begin{array}{cccc}
1&0&0&0\\
0&1&0&0\\
0&0&-1&0\\
0&0&0&1\\
\end{array}\right),
\end{eqnarray}
in the basis
$\{|R\rangle|+\rangle,|R\rangle|-\rangle,|L\rangle|+\rangle,|L\rangle|-\rangle\}$.

Next, the photon passes through a half-wave plate HWP whose optical
axes is set at 22.5$^\circ$ to complete the Hadamard gate ($H^{ph}$)
on the polarization photon,
\begin{eqnarray}                     \label{eq11}
|R\rangle&\xrightarrow{H^{ph}}&|F\rangle\equiv\frac{1}{\sqrt2}(|R\rangle+|L\rangle),\nonumber\\
|L\rangle&\xrightarrow{H^{ph}}&|S\rangle\equiv\frac{1}{\sqrt2}(|R\rangle-|L\rangle).
\end{eqnarray}
That is, after an $H^{ph}$, the state of the whole system becomes
\begin{eqnarray}                      \label{eq12}
\xrightarrow{H^{ph}}|\Psi_4\rangle&=&(\alpha_{c}|L\rangle_5|+\rangle_{c}+\beta_c|R\rangle_5|-\rangle_{c})\nonumber\\
&&\otimes(\alpha_t|+\rangle_{t}+\beta_t|-\rangle_{t}).
\end{eqnarray}
PBS$_3$ transforms the wave packet $|L\rangle_5$ into $|L\rangle_6$,
and transforms $|R\rangle_5$ into $|R\rangle_7$ for interacting with
NV$_2$ and then it reaches PBS$_4$ simultaneously with
$|L\rangle_6$. Before and after the photon passes though NV$_2$, a
Hadamard operation $H^{el}$ is performed on NV$_2$, respectively.
According to Eq. (\ref{eq10}), one can see that the above operations
($H^{el}\rightarrow \text{PBS}_{3}\rightarrow
\text{NV}_{2}\rightarrow \text{PBS}_{4}\rightarrow  H^{el}$)
complete the transformation as
\begin{eqnarray}                      \label{eq13}
\rightarrow
|\Psi_5\rangle&=&\alpha_c\alpha_t|L\rangle_9|+\rangle_{c}|+\rangle_{t}+\alpha_c\beta_t|L\rangle_9|+\rangle_{c}|-\rangle_{t}
\nonumber\\&&
 +\beta_c\alpha_t|R\rangle_9|-\rangle_{c}|-\rangle_{t}+\beta_c\beta_t|R\rangle_9|-\rangle_{c}|+\rangle_{t}.\;\;\;\;\;\;
\end{eqnarray}
Here Hadamard  operation $H^{el}$ completes the following
transformations:
\begin{eqnarray}                     \label{eq14}
|+\rangle &\xrightarrow{H^{el}}& |\vdash\rangle\equiv\frac{1}{\sqrt2}(|+\rangle+|-\rangle),\nonumber\\
|-\rangle
&\xrightarrow{H^{el}}&|\dashv\rangle\equiv\frac{1}{\sqrt2}(|+\rangle-|-\rangle),
\end{eqnarray}

From  Eq. (\ref{eq13}), one can see that to complete the CNOT gate
on two NV centers,  which implements the transformation
\begin{eqnarray}                         \label{eq15}
|\Psi\rangle_{ct} &=&|\psi\rangle_{c}^{e}\otimes|\psi\rangle_{t}^{e} \nonumber\\
                  &=&\alpha_c|+\rangle_{c}(\alpha_t|+\rangle_{t}+\beta_t|-\rangle_{t})\nonumber\\&&
                     +\beta_c|-\rangle_{c}(\alpha_t|+\rangle_{t}+\beta_t|-\rangle_{t}),\nonumber\\
&\xrightarrow{\text{CNOT}}&
                     \alpha_c|+\rangle_{c}(\alpha_t|+\rangle_{t}+\beta_t|-\rangle_{t})\nonumber\\&&
                     +\beta_c|-\rangle_{c}(\alpha_t|-\rangle_{t}+\beta_t|+\rangle_{t}),
\end{eqnarray}
after the photon is detected by the detector $D^F$ or $D^S$ in the
basis
$\{|F\rangle=(|R\rangle+|L\rangle)/\sqrt{2},\;|S\rangle=(|R\rangle-|L\rangle)/\sqrt{2}\}$,
some proper single-qubit operations shown in Tab. \ref{table1}
should be performed on the control qubit and the target qubit,
respectively. Therefore, the quantum circuit  shown in Fig.
\ref{CNOT} performs the CNOT gate on two NV centers, which flips the
state of the target electron qubit in NV$_2$ if and only if (iff)
the control electron qubit in NV$_1$ is in the state $|-\rangle$.
This gate works with a success probability of 100\% in principle.

\begin{table}[htb]
\centering \caption{The feed-forward single unitary operations
performed on the control and the target qubits correspond to the
outcomes of the medium photon for completing the CNOT gate on the
two NV centers with a success probability of 100\%.
$-\sigma_z=-|+\rangle\langle+|+|-\rangle\langle-|$. $I_2$ is a 2
$\times$ 2 unit operation which means doing nothing on a qubit. }

\begin{tabular}{ccc}
\hline  \hline

           & \multicolumn {2}{c}{Feed-forward} \\
\cline{2-3}
   photon              &  $\;\;\;\;$  control qubit   $\;\;\;\;$      &           target qubit          \\
   \hline

$D^F\;(|F\rangle)$   & $I_2$         & $I_2$  \\ 

$D^S\;(|S\rangle)$   &  $-\sigma_z$   & $I_2$       \\  
                             \hline  \hline
\end{tabular}\label{table1}
\end{table}

\section{Solid-state Toffoli gate on a three-qubit NV-center system} \label{sec3}

\begin{figure*}[tpb] 
\begin{center}
\includegraphics[width=16.2 cm,angle=0]{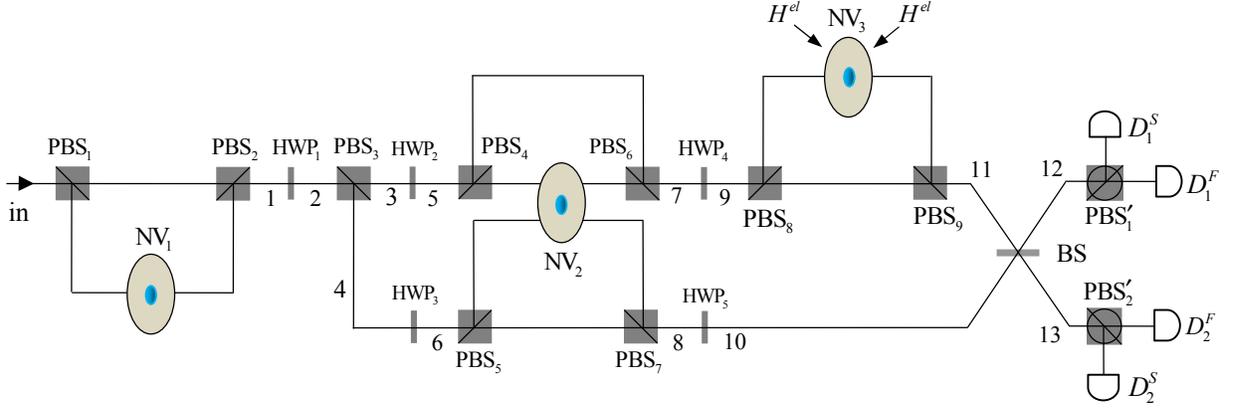}
\caption{(Color online) Compact quantum circuit for
deterministically implementing a Toffoli gate on a quantum system
composed of three NV centers.} \label{Toffoli}
\end{center}
\end{figure*}

A Toffoli gate is used to complete a NOT operation on the state of
the target qubit when both two control qubits are in the state
$|-\rangle$; otherwise, nothing is done on the target qubit. The
principle for implementing a Toffoli gate on a three-qubit NV-center
system is shown in Fig. \ref{Toffoli}. Suppose the first control
qubit $c_1$ in the defect center NV$_1$, the second control qubit
$c_2$ in the defect center NV$_2$, and the target qubit $t$ in the
defect center NV$_3$ are prepared in three arbitrary superposition
electron-spin states as follows:
\begin{eqnarray}                      \label{eq16}
|\psi\rangle_{c_1}^e&=&\alpha_{c_1}|+\rangle_{c_1}+\beta_{c_1}|-\rangle_{c_1},\nonumber\\
|\psi\rangle_{c_2}^e&=&\alpha_{c_2}|+\rangle_{c_2}+\beta_{c_2}|-\rangle_{c_2},\nonumber\\
|\psi\rangle_{t}^e\;&=&\alpha_{t}|+\rangle_{t}+\beta_{t}|-\rangle_{t}.
\end{eqnarray}
Here,
$|\alpha_{c_1}|^2+|\beta_{c_1}|^2=|\alpha_{c_2}|^2+|\beta_{c_2}|^2=|\alpha_{t}|^2+|\beta_{t}|^2=1$.

In order to describe the principle of our Toffoli gate on a
three-qubit NV-center system explicitly,  we specify the evolution
of the system as follows.

An input single-photon medium in the equal polarization
superposition state
$|\psi\rangle^{ph}=(|R\rangle+|L\rangle)/\sqrt{2}$ passes though a
balanced MZ interferometer composed of PBS$_1$, NV$_1$, and PBS$_2$
described by Eq. (\ref{eq10}), and then an $H^{ph}$ (with
$\text{HWP}_1$) is performed on it. PBS$_3$ transforms $|R\rangle_2$
into $|R\rangle_3$, and transforms $|L\rangle_2$ into $|L\rangle_4$.
The evolution of the total states induced by the above operations
($\text{PBS}_1 \rightarrow \text{NV}_1 \rightarrow \text{PBS}_2
\rightarrow \text{HWP}_1 \rightarrow \text{PBS}_3$) can be described
as follows:
\begin{eqnarray}                      \label{eq17}
&&|\Xi_0\rangle=|\psi\rangle^{ph}\otimes|\psi\rangle_{c_1}^{el}\otimes|\psi\rangle_{c_2}^{el}\otimes|\psi\rangle_{t}^{el}
\xrightarrow{\text{PBS}_1,\;\text{NV}_1,\;\text{PBS}_2}\nonumber\\&&
|\Xi_1\rangle=\frac{1}{\sqrt{2}}|R\rangle_1(\alpha_{c_1}|+\rangle_{c_1}+\beta_{c_1}|-\rangle_{c_1})
\otimes|\psi\rangle_{c_2}^e\otimes|\psi\rangle_{t}^e\nonumber\\&&\qquad\quad
              +\frac{1}{\sqrt{2}}|L\rangle_1(-\alpha_{c_1}|+\rangle_{c_1}+ \beta_{c_1}|-\rangle_{c_1})
              \otimes|\psi\rangle_{c_2}^e\otimes|\psi\rangle_{t}^e\nonumber\\
&&\xrightarrow{\text{HWP}_1}
|\Xi_2\rangle=(\alpha_{c_1}|L\rangle_2|+\rangle_{c_1}+\beta_{c_1}|R\rangle_2|-\rangle_{c_1})
\otimes|\psi\rangle_{c_2}^e\nonumber\\&&\qquad\quad\otimes|\psi\rangle_{t}^e\nonumber\\
&&\xrightarrow{\text{PBS}_3}|\Xi_3\rangle=(\alpha_{c_1}|L\rangle_4|+\rangle_{c_1}+\beta_{c_1}|R\rangle_3|-\rangle_{c_1})
\otimes|\psi\rangle_{c_2}^e\nonumber\\&&\qquad\quad\otimes|\psi\rangle_{t}^e.
\end{eqnarray}
Before and after the photon emitting from spatial model 6 (5) passes
through a balanced MZ interferometer composed of PBS$_5$, NV$_2$,
and PBS$_7$ (PBS$_4$, NV$_2$, and PBS$_6$),  an $H^{ph}$ is
performed on it, respectively. These processes ($\text{HWP}_3
\rightarrow \text{PBS}_5 \rightarrow \text{NV}_2 \rightarrow
\text{PBS}_7 \rightarrow \text{HWP}_5$ and $\text{HWP}_2 \rightarrow
\text{PBS}_4 \rightarrow \text{NV}_2 \rightarrow \text{PBS}_6
\rightarrow \text{HWP}_4$) complete the transformation
$|\Xi_3\rangle \rightarrow |\Xi_4\rangle$. Here
\begin{eqnarray}                      \label{eq18}
|\Xi_4\rangle &=&\alpha_{c_1}|+\rangle_{c_1}
(\alpha_{c_2}|R\rangle_{10}|+\rangle_{c_2}+
\beta_{c_2}|L\rangle_{10}|-\rangle_{c_2})
\otimes|\psi\rangle_{t}^e\nonumber\\
                &&+\beta_{c_1}|-\rangle_{c_1} (\alpha_{c_2}|R\rangle_{9}|+\rangle_{c_2}- \beta_{c_2}|L\rangle_{9}|-\rangle_{c_2})
                \otimes|\psi\rangle_{t}^e.\nonumber\\
\end{eqnarray}
The transformation of $\text{PBS}_5 \rightarrow \text{NV}_2
\rightarrow \text{PBS}_7$ can be described by Eq. (\ref{eq10}), and
$\text{PBS}_4 \rightarrow \text{NV}_2 \rightarrow \text{PBS}_6$ can
be written as\begin{eqnarray}                       \label{eq19}
\text{PBS}_{4}\rightarrow \text{NV}_{2}\rightarrow
\text{PBS}_{6}=\left(\begin{array}{cccc}
1&0&0&0\\
0&-1&0&0\\
0&0&1&0\\
0&0&0&1\\
\end{array}\right),
\end{eqnarray}
in the basis
$\{|R\rangle|+\rangle,|R\rangle|-\rangle,|L\rangle|+\rangle,|L\rangle|-\rangle\}$.
 When the photon emits from spatial mode 10, it reaches the
50:50 BS directly. When the photon emits from spatial mode 9, before
it reaches the 50:50 BS, it passes through a balanced MZ
interferometer composed of PBS$_8$, NV$_3$, and PBS$_9$ described by
Eq. (\ref{eq10}), and an $H^{el}$ is performed on the defect NV$_3$
before and after the photon transmits through it, respectively. The
above operations ($H^{el}\rightarrow \text{PBS}_8\rightarrow
\text{NV}_3\rightarrow \text{PBS}_9\rightarrow H^{el}$) complete the
transformation as
\begin{eqnarray}                      \label{eq20}
\rightarrow|\Xi_5\rangle&=&
   \alpha_{c_1}\alpha_{c_2}|R\rangle_{10}|+\rangle_{c_1}|+\rangle_{c_2}(\alpha_t|+\rangle_t+\beta_t|-\rangle_t)\nonumber\\
&&+\alpha_{c_1}\beta_{c_2}|L\rangle_{10}|+\rangle_{c_1}|-\rangle_{c_2}(\alpha_t|+\rangle_t+\beta_t|-\rangle_t)\nonumber\\
&&+\beta_{c_1}\alpha_{c_2}|R\rangle_{11}|-\rangle_{c_1}|+\rangle_{c_2}(\alpha_t|+\rangle_t+\beta_t|-\rangle_t)\nonumber\\
&&+\beta_{c_1}\beta_{c_2}|L\rangle_{11}|-\rangle_{c_1}|-\rangle_{c_2}(\alpha_t|-\rangle_t+\beta_t|+\rangle_t).\;\;\;\;\;\;\;
\end{eqnarray}

Next, the wave packet emitting from spatial 11 interferes with the
wave packet emitting from spatial 10 at the BS, which implements the
transformations
\begin{eqnarray}                     \label{eq21}
|R\rangle_{11}&\xrightarrow{\text{BS}}&\frac{1}{\sqrt2}(|R\rangle_{12}+|R\rangle_{13}),\nonumber\\
|L\rangle_{11}&\xrightarrow{\text{BS}}&\frac{1}{\sqrt2}(|L\rangle_{12}+|L\rangle_{13}),\nonumber\\
|R\rangle_{10}&\xrightarrow{\text{BS}}&\frac{1}{\sqrt2}(|R\rangle_{12}-|R\rangle_{13}),\nonumber\\
|L\rangle_{10}&\xrightarrow{\text{BS}}&\frac{1}{\sqrt2}(|L\rangle_{12}-|L\rangle_{13}).
\end{eqnarray}
$|\Xi\rangle_{5}$ will be transformed into the state
\begin{widetext}
\begin{center}
\begin{eqnarray}                      \label{eq22}
&\xrightarrow{\text{BS}}&|\Xi_{6}\rangle=
\frac{|F\rangle_{12}}{2}\Big[\alpha_{c_1}\alpha_{c_2}|+\rangle_{c_1}|+\rangle_{c_2}(\alpha_t|+\rangle_t+\beta_t|-\rangle_t)
                             +\alpha_{c_1}\beta_{c_2}|+\rangle_{c_1}|-\rangle_{c_2}(\alpha_t|+\rangle_t+\beta_t|-\rangle_t)\nonumber\\&&
                             +\beta_{c_1}\alpha_{c_2}|-\rangle_{c_1}|+\rangle_{c_2}(\alpha_t|+\rangle_t+\beta_t|-\rangle_t)
                             +\beta_{c_1}\beta_{c_2}|-\rangle_{c_1}|-\rangle_{c_2}(\alpha_t|-\rangle_t+\beta_t|+\rangle_t)\Big]\nonumber\\
&&
+\frac{|S\rangle_{12}}{2}\Big[\alpha_{c_1}\alpha_{c_2}|+\rangle_{c_1}|+\rangle_{c_2}(\alpha_t|+\rangle_t+\beta_t|-\rangle_t)
                             -\alpha_{c_1}\beta_{c_2}|+\rangle_{c_1}|-\rangle_{c_2}(\alpha_t|+\rangle_t+\beta_t|-\rangle_t)\nonumber\\&&
                             +\beta_{c_1}\alpha_{c_2}|-\rangle_{c_1}|+\rangle_{c_2}(\alpha_t|+\rangle_t+\beta_t|-\rangle_t)
                             -\beta_{c_1}\beta_{c_2}|-\rangle_{c_1}|-\rangle_{c_2}(\alpha_t|-\rangle_t+\beta_t|+\rangle_t)\Big]\nonumber\\
&&
+\frac{|F\rangle_{13}}{2}\Big[-\alpha_{c_1}\alpha_{c_2}|+\rangle_{c_1}|+\rangle_{c_2}(\alpha_t|+\rangle_t+\beta_t|-\rangle_t)
                             -\alpha_{c_1}\beta_{c_2}|+\rangle_{c_1}|-\rangle_{c_2}(\alpha_t|+\rangle_t+\beta_t|-\rangle_t)\nonumber\\&&
                             +\beta_{c_1}\alpha_{c_2}|-\rangle_{c_1}|+\rangle_{c_2}(\alpha_t|+\rangle_t+\beta_t|-\rangle_t)
                             +\beta_{c_1}\beta_{c_2}|-\rangle_{c_1}|-\rangle_{c_2}(\alpha_t|-\rangle_t+\beta_t|+\rangle_t)\Big]\nonumber\\
&&
+\frac{|S\rangle_{13}}{2}\Big[-\alpha_{c_1}\alpha_{c_2}|+\rangle_{c_1}|+\rangle_{c_2}(\alpha_t|+\rangle_t+\beta_t|-\rangle_t)
                             +\alpha_{c_1}\beta_{c_2}|+\rangle_{c_1}|-\rangle_{c_2}(\alpha_t|+\rangle_t+\beta_t|-\rangle_t)\nonumber\\&&
                             +\beta_{c_1}\alpha_{c_2}|-\rangle_{c_1}|+\rangle_{c_2}(\alpha_t|+\rangle_t+\beta_t|-\rangle_t)
                             -\beta_{c_1}\beta_{c_2}|-\rangle_{c_1}|-\rangle_{c_2}(\alpha_t|-\rangle_t+\beta_t|+\rangle_t)\Big].
\end{eqnarray}
\end{center}
\end{widetext}

The photon medium is measured in the basis $\{|F\rangle,|S\rangle\}$
by the detector $D_i^F$ or $D_i^S$.  Following with the feedforward
operations performed on the NV centers, shown in Table
\ref{tableToffoli}, we accomplish  the construction of the Toffoli
gate on the three NV centers in a deterministic way. That is, the
state of the system composed of the three defect NV$_1$, NV$_2$, and
NV$_3$ becomes
\begin{eqnarray}                      \label{eq23}
|\Xi\rangle_{\text{Toffoli}}&=&
\alpha_{c_1}\alpha_{c_2}|+\rangle_{c_1}|+\rangle_{c_2}(\alpha_t|+\rangle_t+\beta_t|-\rangle_t)\nonumber\\&&
+\alpha_{c_1}\beta_{c_2}|+\rangle_{c_1}|-\rangle_{c_2}(\alpha_t|+\rangle_t+\beta_t|-\rangle_t)\nonumber\\&&
+\beta_{c_1}\alpha_{c_2}|-\rangle_{c_1}|+\rangle_{c_2}(\alpha_t|+\rangle_t+\beta_t|-\rangle_t)\nonumber\\&&
+\beta_{c_1}\beta_{c_2}|-\rangle_{c_1}|-\rangle_{c_2}(\alpha_t|-\rangle_t+\beta_t|+\rangle_t).\;\;\;\;\;\;
\end{eqnarray}

From the processes above, one can see that the setup shown in Fig.
\ref{Toffoli} completes the transformation,
\begin{eqnarray}                      \label{eq24}
&&|\Xi\rangle_{c_1,\;c_2,\;t}=|\psi\rangle_{c_1}^{el}\otimes|\psi\rangle_{c_2}^{el}\otimes|\psi\rangle_{t}^{el}\nonumber\\
&\xrightarrow{\text{Toffoli}}&
\alpha_{c_1}\alpha_{c_2}|+\rangle_{c_1}|+\rangle_{c_2}(\alpha_t|+\rangle_t+\beta_t|-\rangle_t)\nonumber\\&&
                             +\alpha_{c_1}\beta_{c_2}|+\rangle_{c_1}|-\rangle_{c_2}(\alpha_t|+\rangle_t+\beta_t|-\rangle_t)\nonumber\\&&
                             +\beta_{c_1}\alpha_{c_2}|-\rangle_{c_1}|+\rangle_{c_2}(\alpha_t|+\rangle_t+\beta_t|-\rangle_t) \nonumber\\&&
                             +\beta_{c_1}\beta_{c_2}|-\rangle_{c_1}|-\rangle_{c_2}(\alpha_t|-\rangle_t+\beta_t|+\rangle_t).\;\;\;\;\;\;
\end{eqnarray}
That is, the setup shown in Fig. \ref{Toffoli} realizes exactly the
Toffoli gate on the three-qubit NV-center system, which flips the
state of the target qubit iff  both the two control qubits are  in
the state $|-\rangle$.

\begin{table}[htb]
\centering \caption{The operations performed on the control and the
target qubits correspond to the measurement outcomes of the medium
photon for completing the Toffoli gate on the three NV centers with
a success probability of 100\%.}
\begin{tabular}{cccc}
\hline  \hline

           & \multicolumn {3}{c}{Feedforward} \\
\cline{2-4}
photon                         &$\;\;$  qubit $c_1$   $\;\;$     &  $\;\;$  qubit $c_2$    $\;\;$     &  qubit $t$ \\
\hline
$D_1^F\; (|F\rangle_{12})$     & $I_2$                &    $I_2$            &   $I_2$   \\
$D_1^S\;(|S\rangle_{12})$      & $I_2$                &    $ \sigma_z$      &  $I_2$   \\
$D_2^F\;(|F\rangle_{13})$      &  $-\sigma_z$         &  $I_2$              &  $I_2$   \\
$D_2^S\;(|S\rangle_{13})$      &  $-\sigma_z$         &  $\sigma_z$         &  $I_2$   \\

                             \hline  \hline
\end{tabular}\label{tableToffoli}
\end{table}

\section{Solid-state Fredkin gate on a three-qubit NV-center system} \label{sec4}

\begin{figure*}[tpb] 
\begin{center}
\includegraphics[width=16.2 cm,angle=0]{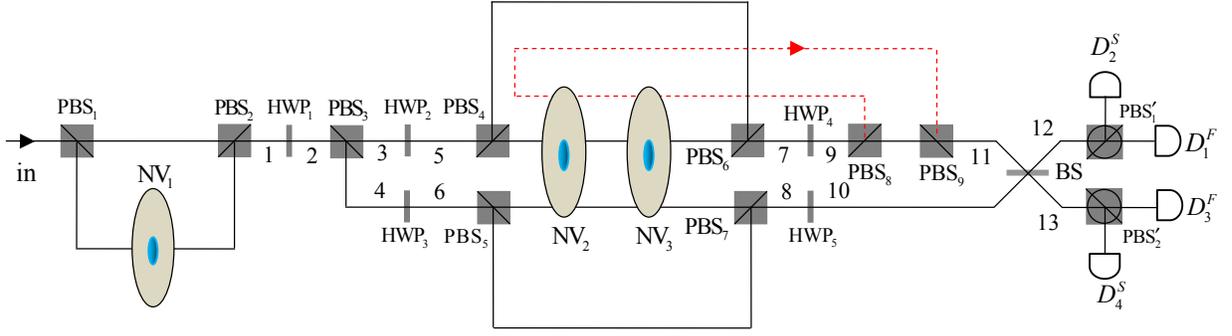}
\caption{(Color online) Schematic setup for deterministically
implementing a Fredkin gate on three NV centers.} \label{Fredkin}
\end{center}
\end{figure*}

A Fredkin gate is used to exchange the states of the two target
qubits iff the control qubit is in the state $|-\rangle$. Our
quantum circuit for implementing a Fredkin gate on a three-qubit
NV-center system in a deterministic way is shown in Fig.
\ref{Fredkin}. The control qubit $c$ encoded on NV center ``NV$_1$",
the first target qubit $t_1$ encoded on NV center ``NV$_2$", and the
second target qubit $t_2$ encoded on NV center ``NV$_3$"  are
initially prepared in three arbitrary states
\begin{eqnarray}                      \label{eq25}
&&|\psi\rangle_{c}^{el}=\alpha_{c}|+\rangle_{c}+\beta_{c}|-\rangle_{c},\nonumber\\
&&|\psi\rangle_{t_1}^{el}=\alpha_{t_1}|+\rangle_{t_1}+\beta_{t_1}|-\rangle_{t_1},\nonumber\\
&&|\psi\rangle_{t_2}^{el}=\alpha_{t_2}|+\rangle_{t_2}+\beta_{t_2}|-\rangle_{t_2}.
\end{eqnarray}
Here
$|\alpha_{c}|^2+|\beta_{c}|^2=|\alpha_{t_1}|^2+|\beta_{t_1}|^2=|\alpha_{t_2}|^2+|\beta_{t_2}|^2=1$.
The photon medium $p$ is prepared in the equal superposition state
\begin{eqnarray}                      \label{eq26}
|\psi\rangle^{ph}=\frac{1}{\sqrt{2}}(|R\rangle+|L\rangle).
\end{eqnarray}
That is, the initial state of the quantum  system, composed of the
three electrons $c$, $t_1$, and $t_2$, and a single photon $p$, can
be written as
\begin{eqnarray}                      \label{eq27}
|\Pi_0\rangle=|\psi\rangle^{ph}
\otimes|\psi\rangle_{c}^{el}\otimes|\psi\rangle_{t_1}^{el}
\otimes|\psi\rangle_{t_2}^{el}.
\end{eqnarray}

In the following, let us discuss the construction of the solid-state
Fredkin gate on a three-qubit NV-center system step by step.

First, a photon medium is injected into the input port $in$ and it
passes through a balanced MZ interferometer composed of PBS$_1$,
NV$_1$, and PBS$_2$,  and then an $H^{ph}$ is performed on it (i.e.,
let it pass through HWP$_1$). PBS$_3$ transmits the $R$-polarized
photon to spatial model 3, and reflects the $L$-polarized photon to
spatial model 4. Based on the argument as made in Sec. \ref{sec3},
one can see that the state of the whole system composed of a single
photon medium and three NV centers then becomes
\begin{eqnarray}                      \label{eq28}
|\Pi_1\rangle=(\alpha_{c}|L\rangle_4|+\rangle_{c}+\beta_{c}|R\rangle_3|-\rangle_{c})\otimes|\psi\rangle_{t_1}^e\otimes|\psi\rangle_{t_2}^e.
\end{eqnarray}
Before and after the photon emitting from spatial model 6 (5) passes
through a balanced MZ interferometer composed of PBS$_5$, NV$_2$,
NV$_3$, and PBS$_7$ (PBS$_4$, NV$_2$, NV$_3$, and PBS$_6$),  an
$H^{ph}$ is performed on it, respectively. The  state of the
complicated system  after these operations ($\text{HWP}_3
\rightarrow \text{PBS}_5 \rightarrow \text{NV}_2 \rightarrow
\text{NV}_3 \rightarrow \text{PBS}_7 \rightarrow \text{HWP}_5$ and
$\text{HWP}_2 \rightarrow \text{PBS}_4 \rightarrow \text{NV}_2
\rightarrow \text{NV}_3 \rightarrow \text{PBS}_6 \rightarrow
\text{HWP}_4$) becomes
\begin{eqnarray}                      \label{eq29}
\rightarrow  |\Pi_2\rangle &=&
\alpha_{c}\alpha_{t_1}\alpha_{t_2}|L\rangle_{10}|+\rangle_{c}|+\rangle_{t_1}|+\rangle_{t_2}\nonumber\\
&&
-\alpha_{c}\alpha_{t_1}\beta_{t_2}|R\rangle_{10}|+\rangle_{c}|+\rangle_{t_1}|-\rangle_{t_2}\nonumber\\
&&
-\alpha_{c}\beta_{t_1}\alpha_{t_2}|R\rangle_{10}|+\rangle_{c}|-\rangle_{t_1}|+\rangle_{t_2}\nonumber\\
&&
+\alpha_{c}\beta_{t_1}\beta_{t_2}|L\rangle_{10}|+\rangle_{c}|-\rangle_{t_1}|-\rangle_{t_2}\nonumber\\
&&
+\beta_{c}\alpha_{t_1}\alpha_{t_2}|R\rangle_{9}|-\rangle_{c}|+\rangle_{t_1}|+\rangle_{t_2}\nonumber\\
&&
-\beta_{c}\alpha_{t_1}\beta_{t_2}|L\rangle_{9}|-\rangle_{c}|+\rangle_{t_1}|-\rangle_{t_2}\nonumber\\
&&
-\beta_{c}\beta_{t_1}\alpha_{t_2}|L\rangle_{9}|-\rangle_{c}|-\rangle_{t_1}|+\rangle_{t_2}\nonumber\\
&&
+\beta_{c}\beta_{t_1}\beta_{t_2}|R\rangle_{9}|-\rangle_{c}|-\rangle_{t_1}|-\rangle_{t_2}.
\end{eqnarray}
Here the balanced MZ interferometer composed of PBS$_5$, NV$_2$,
NV$_3$, and PBS$_7$ (PBS$_4$, NV$_2$, NV$_3$, and PBS$_6$) completes
the unitary operation
\begin{eqnarray}                       \label{eq30}
&&\text{PBS}_{5(4)}
\rightarrow\text{NV}_{2}\rightarrow\text{NV}_{3}\rightarrow\text{PBS}_{7(6)}\nonumber\\&&\qquad\qquad\qquad\qquad\qquad\quad
=\left(\begin{array}{cccc}
1&0&0&0\\
0&-1&0&0\\
0&0&-1&0\\
0&0&0&I_5\\
\end{array}\right),\;\;\;\;\;\;
\end{eqnarray}
in the basis $\{|R\rangle|+\rangle|+\rangle,
|R\rangle|+\rangle|-\rangle, |R\rangle|-\rangle|+\rangle,
|R\rangle|-\rangle|-\rangle,\\ |L\rangle|+\rangle|+\rangle,
|L\rangle|+\rangle|-\rangle, |L\rangle|-\rangle|+\rangle,
|L\rangle|-\rangle|-\rangle\}$.

Next, when the photon emits from spatial mode 10, it reaches the
50:50 BS directly. When the photon emits from spatial mode 9, before
it reaches the BS, it passes through a balanced MZ interferometer
composed of PBS$_8$, PBS$_9$, NV$_2$ and NV$_3$, which completes the
operation
\begin{eqnarray}                       \label{eq31}
\text{PBS}_{8} \rightarrow \text{NV}_{3} \rightarrow \text{NV}_{2}
\rightarrow \text{PBS}_{9}=\left(\begin{array}{ccccc}
I_5&0&0&0\\
0&-1&0&0\\
0&0&-1&0\\
0&0&0&1\\
\end{array}\right).\;\;\;\;
\end{eqnarray}
Before and after the photon interacts with NV$_3$ and NV$_2$, an
$H^{el}$ is performed on NV$_3$ and NV$_2$, respectively. These
operations
($H^{el}\rightarrow\text{PBS}_8\rightarrow\text{NV}_3\rightarrow\text{NV}_2\rightarrow\text{PBS}_9\rightarrow
H^{el}$) complete the transformation
$|\Pi_2\rangle\rightarrow|\Pi_3\rangle$. Here
\begin{eqnarray}                      \label{eq32}
 |\Pi_3\rangle&=&
\alpha_{c}\alpha_{t_1}\alpha_{t_2}|L\rangle_{10}|+\rangle_{c}|+\rangle_{t_1}|+\rangle_{t_2}\nonumber\\&&
-\alpha_{c}\alpha_{t_1}\beta_{t_2}|R\rangle_{10}|+\rangle_{c}|+\rangle_{t_1}|-\rangle_{t_2}\nonumber\\&&
-\alpha_{c}\beta_{t_1}\alpha_{t_2}|R\rangle_{10}|+\rangle_{c}|-\rangle_{t_1}|+\rangle_{t_2}\nonumber\\&&
+\alpha_{c}\beta_{t_1}\beta_{t_2}|L\rangle_{10}|+\rangle_{c}|-\rangle_{t_1}|-\rangle_{t_2}\nonumber\\&&
+\beta_{c}\alpha_{t_1}\alpha_{t_2}|R\rangle_{11}|-\rangle_{c}|+\rangle_{t_1}|+\rangle_{t_2}\nonumber\\&&
-\beta_{c}\alpha_{t_1}\beta_{t_2}|L\rangle_{11}|-\rangle_{c}|-\rangle_{t_1}|+\rangle_{t_2}\nonumber\\&&
-\beta_{c}\beta_{t_1}\alpha_{t_2}|L\rangle_{11}|-\rangle_{c}|+\rangle_{t_1}|-\rangle_{t_2}\nonumber\\&&
+\beta_{c}\beta_{t_1}\beta_{t_2}|R\rangle_{11}|-\rangle_{c}|-\rangle_{t_1}|-\rangle_{t_2}.
\end{eqnarray}
The 50:50 BS, described by Eq. (\ref{eq21}), transforms
$|\Pi_3\rangle$ into
\begin{eqnarray}                      \label{eq33}
|\Pi_{4}\rangle&=& \frac{|F_{12}\rangle}{2}
\Big[\alpha_{c}|+\rangle_{c}(\alpha_{t_1}|+\rangle_{t_1}-\beta_{t_1}|-\rangle_{t_1})\nonumber\\&&\times(\alpha_{t_2}|
+\rangle_{t_2}-\beta_{t_2}|-\rangle_{t_2})
    +\beta_{c}|-\rangle_{c}(\alpha_{t_2}|+\rangle_{t_1}\nonumber\\&&-\beta_{t_2}|-\rangle_{t_1})(\alpha_{t_1}|
    +\rangle_{t_2}-\beta_{t_1}|-\rangle_{t_2})\Big]\nonumber\\&&
+\frac{|S_{12}\rangle}{2}
\Big[-\alpha_{c}|+\rangle_{c}(\alpha_{t_1}|+\rangle_{t_1}+\beta_{t_1}|-\rangle_{t_1})\nonumber\\&&\times(\alpha_{t_2}|
+\rangle_{t_2}+\beta_{t_2}|-\rangle_{t_2})
    +\beta_{c}|-\rangle_{c}(\alpha_{t_2}|+\rangle_{t_1}\nonumber\\&&+\beta_{t_2}|-\rangle_{t_1})(\alpha_{t_1}|+\rangle_{t_2}
    +\beta_{t_1}|-\rangle_{t_2})\Big]\nonumber\\&&
+\frac{|F_{13}\rangle}{2}
\Big[-\alpha_{c}|+\rangle_{c}(\alpha_{t_1}|+\rangle_{t_1}-\beta_{t_1}|-\rangle_{t_1})\nonumber\\&&\times(\alpha_{t_2}|
+\rangle_{t_2}-\beta_{t_2}|-\rangle_{t_2})
    +\beta_{c}|-\rangle_{c}(\alpha_{t_2}|+\rangle_{t_1}\nonumber\\&&-\beta_{t_2}|-\rangle_{t_1})(\alpha_{t_1}|
    +\rangle_{t_2}-\beta_{t_1}|-\rangle_{t_2})\Big]\nonumber\\&&
+\frac{|S_{13}\rangle}{2}
\Big[\alpha_{c}|+\rangle_{c}(\alpha_{t_1}|+\rangle_{t_1}+\beta_{t_1}|-\rangle_{t_1})\nonumber\\&&\times(\alpha_{t_2}|
+\rangle_{t_2}+\beta_{t_2}|-\rangle_{t_2})
    +\beta_{c}|-\rangle_{c}(\alpha_{t_2}|+\rangle_{t_1}\nonumber\\&&+\beta_{t_2}|-\rangle_{t_1})(\alpha_{t_1}|
    +\rangle_{t_2}+\beta_{t_1}|-\rangle_{t_2})\Big].
\end{eqnarray}

Third, by detecting the single-photon medium in the basis
$\{|F\rangle,|S\rangle\}$ and following with the feedforward
single-qubit unitary operations shown in Table \ref{tableFredkin},
one can see that the state of the system composed of NV$_1$, NV$_2$,
and NV$_3$ becomes
\begin{eqnarray}                      \label{eq34}
|\Pi\rangle_{\text{Fredkin}}&=&
\alpha_{c}|+\rangle_{c}(\alpha_{t_1}|+\rangle_{t_1}+\beta_{t_1}|-\rangle_{t_1})\nonumber\\&&\times(\alpha_{t_2}|+\rangle_{t_2}
+\beta_{t_2}|-\rangle_{t_2})\nonumber\\&&
    +\beta_{c}|-\rangle_{c}(\alpha_{t_2}|+\rangle_{t_1}+\beta_{t_2}|-\rangle_{t_1})\nonumber\\&&\times(\alpha_{t_1}|+\rangle_{t_2}
    +\beta_{t_1}|-\rangle_{t_2}).
\end{eqnarray}

Comparing  Eq. (\ref{eq28}) with Eq. (\ref{eq34}), one can see that
the quantum circuit shown in Fig. \ref{Fredkin} implements a Fredkin
gate on the three NV centers with the success probability of 100\%
in principle, which swaps the states of two target qubits iff the
control qubit is in the state $|-\rangle$.

\begin{table}[!h]
\centering \caption{The operations performed on the control and the
target qubits correspond to the measurement outcomes of the medium
photon for completing the Fredkin gate on the three NV centers with
a success probability of 100\%.}
\begin{tabular}{cccc}
\hline  \hline

           & \multicolumn {3}{c}{Feedforward} \\
\cline{2-4}
photon                        & $\;\;$ qubit $c$ $\;\;$     &  $\;\;$ qubit $t_1$    $\;\;$   &   $\;\;$ qubit $t_2$   \\
\hline
$D_1^F\;(|F\rangle_{12})$            &   $I_2$              &   $\sigma_z$               &    $\sigma_z$           \\
$D_1^S\;(|S\rangle_{12})$            &   $-\sigma_z$        &   $I_2$                    &    $I_2$                \\
$D_2^F\;(|F\rangle_{13})$            &   $-\sigma_z$        &   $\sigma_z$               &    $\sigma_z$           \\
$D_2^S(\;|S\rangle_{13})$            &   $I_2$              &   $I_2$                    &    $I_2$                \\
                             \hline  \hline
\end{tabular}\label{tableFredkin}
\end{table}

\section{Fidelities and efficiencies of our universal quantum gates} \label{sec5}

Let us estimate the fidelities and the efficiencies of our universal
solid-state quantum  gates  discussed  above, defining the fidelity
as $F=|\langle\psi_{\text{real}}|\psi_{\text{ideal}}\rangle|^2$.
Here, $|\psi_{\text{ideal}}\rangle$ is the target state of the
NV-center-cavity system  encoded for the quantum gate in the ideal
case $g\geq5\sqrt{\kappa\gamma}$, and $|\psi_{\text{real}}\rangle$
is the target state of a realistic NV-center-cavity system. Defining
the efficiency as the yield of the photons, that is,
$\eta=n_{\text{output}}/n_{\text{input}}$. Here, $n_{\text{input}}$
is the number of the input photon, whereas $n_{\text{output}}$ is
the number of the output photon. The gates are realized by the
input-output processes of the photon medium, which means that the
reflection coefficient of the NV-cavity system determines the
fidelities and the efficiencies of our universal quantum gates.

\begin{figure}[!h]
\begin{center}
\includegraphics[width=7.8 cm,angle=0]{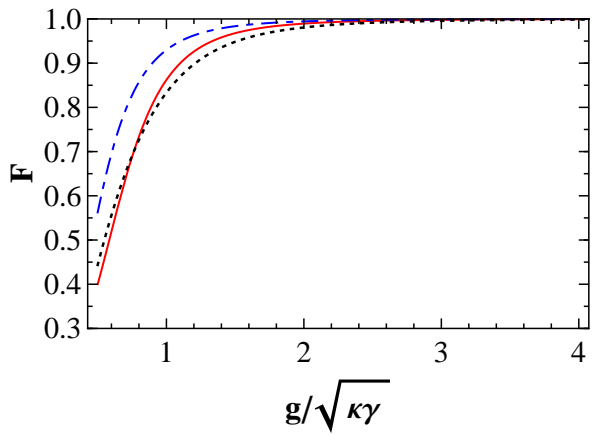}
\caption{(Color online) The fidelities of the CNOT (solid line,
red),  Toffoli (dash-dotted line, blue), and Fredkin (dotted line,
black) gates vs $g/\sqrt{\kappa\gamma}$. Here,
$g/\sqrt{\kappa\gamma}\geq 0.5$.} \label{fidelity}
\end{center}
\end{figure}

\begin{figure}[!h]
\begin{center}
\includegraphics[width=7.8 cm,angle=0]{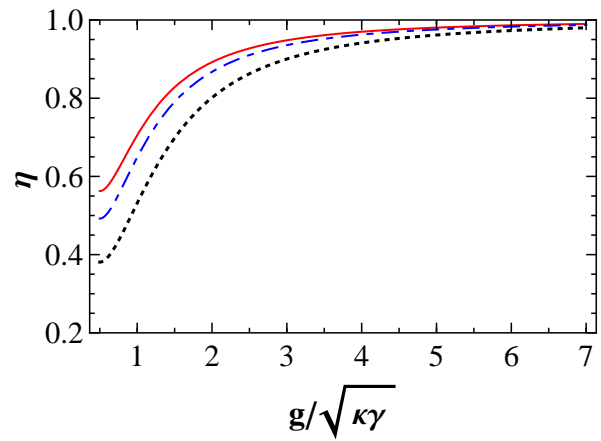}
\caption{(Color online)  The efficiencies of the CNOT (solid line,
red), Toffoli (the dash-dotted line, blue), and Fredkin (dotted
line, black) gates vs $g/\sqrt{\kappa\gamma}$ gates vs
$g/\sqrt{\kappa\gamma}$. Here, $g/\sqrt{\kappa\gamma}\geq 0.5$.}
\label{efficiency}
\end{center}
\end{figure}

Combing the specific evolutions of the CNOT, Toffoli, and Fredkin
gates and the input-output relations of the NV-cavity system in the
realistic case given by Eq. (\ref{eq4}), one can see that the
fidelities of those gates can be calculated as
\begin{eqnarray}       \label{eq35}
&&F_{\text{CNOT}}=\frac{(2+|r|+|r|^2)^2}{2(5 - 2|r| + 2|r|^2 + 2|r|^3 +|r|^4)},\nonumber\\
&&F_{\text{Toffoli}}=\frac{(3+|r|)^4}{16(3+|r|^2)^2},\nonumber\\
&&F_{\text{Fredkin}}=\frac{\zeta_{\text{Fredkin}}}{\xi_{\text{Fredkin}}},
\end{eqnarray}
with
\begin{eqnarray}       \label{eq36}
\zeta_{\text{Fredkin}}&=&(29 + 19 |r| + 8|r|^2 + 4 |r|^3 + 3 |r|^4 + |r|^5)^2,\nonumber\\
\xi_{\text{Fredkin}}&=&8 [237 - 10 |r| + 165 |r|^2 - 8 |r|^3 + 66
|r|^4\nonumber\\&& - 12 |r|^5 + 26 |r|^6 +
   |r|^7 (3 + |r|)\nonumber\\&&\times (8 + 3 |r| + |r|^2)].
\end{eqnarray}

 The efficiencies of those gates can be calculated as
\begin{eqnarray}       \label{eq37}
&&\eta_{\text{CNOT}}=\left[\frac{3+|r|^2}{4}\right]^2,\nonumber\\
&&\eta_{\text{Toffoli}}=\frac{(3+|r|^2)^2(7+|r|^2)}{128},\\
&&\eta_{\text{Fredkin}}=\frac{(3+|r|^2)[4+(1+|r|^2)^2][12+(1+|r|^2)^2]}{512}.\nonumber
\end{eqnarray}

For the diamond NV centers, the photoluminescence is partially
unpolarized, and the emission with ZPL is only 4\% of the total
emission.  $\gamma_{\text{ZPL}}$ with zero phonon line is only 4\%
of $\gamma_{\text{total}}=2\pi\times15$ MHz \cite{photon-NV,split}.
$Q=c/\lambda\kappa$, where $c$ is the speed of light and
$\lambda=637$ $nm$ is the transition wavelength.  The WGM cavities
with  microtoroidal form have   attracted much attention
\cite{microtoroid5}. Ref. \cite{microtoroid5} shows that the
polymer-coated microtoroid is feasible and robust in experiments.
For the diamond NV center in a MTR with WGM mode system, Ref.
\cite{photon-NV} shows that when $g/\sqrt{\kappa\gamma}\geq 3$ with
$\omega_c=\omega_p=\omega_0$, $r(\omega_p)\sim$0.95; when
$g/\sqrt{\kappa\gamma}\geq 5$ with $\omega_c=\omega_p=\omega_0$,
$Q\sim10^5$ (corresponding to $\kappa\sim$ 1 GHz) or $Q\sim10^4$
(corresponding to $\kappa\sim$ 10 GHz), $r(\omega_p)\sim$1.

Figures \ref{fidelity} and  \ref{efficiency} show the fidelities and
the efficiencies of our universal quantum gates  as a function of
$g/\sqrt{\kappa\gamma}$ with $\omega_c=\omega_p=\omega_0$ and
$g/\sqrt{\kappa\gamma}\geq 1/2$. Our results show that the
fidelities and the efficiencies of our quantum gates increase with
$g/\sqrt{\kappa\gamma}$. When $g/\sqrt{\kappa\gamma}=5$, the
fidelities of the gates are unity with $\eta_{\text{CNOT}}=98.05\%$,
$\eta_{\text{Toffoli}}=97.57\%$, and
$\eta_{\text{Fredkin}}=96.15\%$.

\section{Discussion and summary}\label{sec6}

Universal quantum gates in solid-state systems are much more
attractive as they have a good scalability. Many schemes have been
proposed for realizing universal quantum gates on solid-state
systems. Based on superconductor, Romero \emph{et al.}
\cite{superPRL} and Stojanovi\'{c} \emph{et al.} \cite{superPRB}
proposed some interesting schemes for realizing controlled-phase and
Toffoli gates in nanosecond time scale, respectively. Liang and Li
\cite{teleportation} proposed a scheme for realizing a SWAP gate
between the flying and the stationary qubits.  In 2010, the quantum
circuit for realizing a CNOT between a quantum-dot qubit and a
polarized photon qubit was designed by Bonato \emph{et al.}
\cite{CNOT-Hybrid}. Based on appealing diamond NV-center qubits,
Yang \emph{et al.} \cite{CCPF} proposed a scheme for realizing a
conditional phase gate between NV centers assisted by high-Q silica
microsphase cavity, and the control and the target qubits are
encoded on different energy levels. Jelezko \emph{et al.}
\cite{CROT} designed a quantum circuit for realizing controlled-ROT
gate between an electron and a nuclear spin qubits in a NV center.

The schemes we proposed for constructing the two-qubit CNOT, and
three-qubit Toffoli and Fredkin gates on diamond NV centers inside
resonators have some interesting features. (1) The quantum circuits
are compact. Especially the schemes for CNOT and Toffoli gates, in
which the photon medium only interacts with each qubit one time. The
complexity of our schemes for Toffoli and Fredkin gates beats its
synthesis procedure. The optimal synthesis of a Toffoli
\cite{Toffolicost} gates requires six CNOT gates. A Fredkin gate can
be decomposed into six specific gates \cite{Fredkincost}, i.e., two
CNOT and three controlled-$\sqrt{\text{NOT}}$ gates.  (2) Our
schemes are economic. Auxiliary electron qubits are employed in
Refs. \cite{CNOT1,topological}, but they are not required in our
schemes. Furthermore, only one single-photon medium is employed in
our proposals. (3) The static electron qubits employed in our
proposals are more robust than  the moving qubits in Ref.
\cite{CNOT1}. (4) Different from Refs.
\cite{CROT,CNOT-Hybrid,teleportation} (the hybrid qubits are
employed), all the qubits in our proposals are encoded on the spins
of the electrons associated with NV centers, which means our quantum
gates are scalable. Unfortunately, identical NV centers are required
in our proposals, although the identical NV centers are challenge
with current techniques, the energy levels of different NV centers
can be adjusted by external magnetic fields. (5) They have a long
coherence time in NV centers even at the room temperature. (6)
Different from Ref. \cite{CCPF}, all of the qubits in our proposals
are encoded on the identical energy levels.  (7) Our proposals are
robust against low strain and magnetic fields, due to the special
auxiliary energy level we employed. (8) Compared with an atom-cavity
system, the time scale for manipulating an NV center is much shorter
than that seen with an atom. Also it is difficult to trapped an atom
in the cavity. Although high fidelities and efficiencies can be
achieved in our schemes, only 4\% of the emitted photon emitting
from the NV centers are coherent emissions within the narrowband ZPL
at 637 $nm$ due to the particular characteristic of the NV centers.

In summary, we have designed the compact quantum circuits for
implementing some deterministic universal quantum gates on NV
centers, including the CNOT,  Toffoli, and Fredkin gates, by means
of the interaction between an NV-cavity-assisted qubit and a
single-photon medium in a scalable fashion. The quantum gates are
constructed by some input-output processes of a single photon
medium, the measurements on the polarizations of the photon medium,
and feedforward operations. As these quantum gates have a long
coherence time even at the room temperature and they are universal,
intrinsically deterministic, and scalable, they provide a different
way for quantum computing in solid-state quantum systems.

\bigskip

\section*{ACKNOWLEDGEMENTS}

This work is supported by the National Natural Science Foundation of
China under Grant No. 11174039  and  NECT-11-0031

\end{document}